\documentstyle[12pt]{article}

\begin{document}
\centerline{\bf Quantum teleportation of a single-photon wave packet}
\vskip 1mm
\centerline{S.N.Molotkov$^{\#,\dag}$}
\vskip 1mm
\centerline{\sl\small $^{\#}$Institute of Solid State Physics, 
Russian Academy of Sciences}
\centerline{\sl\small 142432, Chernogolovka, Moscow District, Russia}
\centerline{\sl\small $^{\dag}$Wihuri Physical Laboratory, University of 
Turku, 20500 Turku, Finland} 
\vskip 1mm
\begin{abstract}
A quantum teleportation scheme based on the EPR-pair entangled with 
respect to the ``energy+time'' variables is proposed.
Teleportation of the multimode state of a single-photon wave packet 
is considered.
\end{abstract} 
\vskip 1mm
The fundamental unit of quantum information is represented by a quantum bit 
(qubit) [1].
Qubit can be associated with arbitrary two-level quantum system
(e.g. spin-1/2 particle). Any two orthogonal states of the quantum system
can be identified with the boolean values 0 and 1 (states of a classical bit).
Contrary to its classical counterpart (bit), the quantum beat can
be found in an arbitrary superposition of the states 0 and 1.
Different quantum bits can also be found in various entangled states.
A quantum bit whose state is not known beforehand cannot be cloned [2].
There exists a fundamental law which prohibits
cloning of an unknown quantum state [2,3]. Also, no single measurement
can provide comprehensive information on the state of a quantum bit.
However, a qubit can be reliably relayed by making its replica
(quantum teleportation [4]).
A fundamental difference between the clone and the copy is that the 
operator who has produced a replica of an unknown quantum bit
does not know himself the state of the replica which is completely 
identical to the original quantum bit [4].
The quantum teleportation employs non-local quantum correlations
(Einstein-Podolsky-Rosen effect) [5].

To produce a replica of an unknown state of the two-level quantum system
(for example, spin-1/2 particle), user {\it A} first generates an EPR-pair,
i.e. two spin-1/2 particles in an entangled state. Then he leaves one of 
the twins in his own apparatus and sends the other one to user {\it B}. 
After that user {\it A} performs joint measurements of the particle whose
state is unknown to him and his twin from the EPR-pair. The measurements
are made in the so-called Bell basis [6]. The measurements produce four 
equiprobable outcomes which can be reliably distinguished from each other
because of the basis orthogonality (after the measurement the state of
the pair is known). Because of the initial correlations in the EPR-pair,
the particle sent to user {\it B} is rendered to a new state which is 
a perfect replica of the unknown state to within a unitary rotation [4].
The four outcomes of the measurement performed by user {\it A} 
yield two bits of classical information and indicate which unitary 
rotation should be applied by user {\it B} to his particle to produce 
the state identical to the initial unknown state. Reliable teleportation 
of a single qubit requires one EPR-pair (two qubits in an entangled state, 
i.e. one entangled bit (ebit [7])) and two classical bits of information.
 
Quantum teleportation has recently been demonstrated experimentally for
a photon with unknown polarization [8].

The problem of teleportation of the wave function in the one-dimensional
case where momentum and coordinate play the role of continuous dynamical 
variables was analyzed in Ref.[9] where the EPR-pair wave function
was as a matter of fact replaced by a singular wave function from Ref. [5]:
\begin{equation}
\psi(x_1,x_2)\propto \delta(x_1+x_2-X_0)\delta(p_1-p_2),
\end{equation}
where $(x_1,x_2)$ ? $(p_1,p_2)$ correspond to canonically conjugate
variables of the first and second particles in the EPR-pair.
Such a singular wavefunction corresponds the ideal EPR correlations, and
the closer the wavefunction to Eq.(1), the lower is the probability
of obtaining the classical information sent by user A to user B,
although the replica itself in the limit (1) tends to the ideal copy
of the initial unknown state. Investigated in a recent paper [10]
was the teleportation of a quantum state described by the dynamical 
variables $(x,p)$ (the unknown states in Ref.[10] correspond to a single-mode
of the photon) for the case of non-ideal EPR-correlations. Imperfect
EPR correlations reduce the replica accuracy but simultaneously
enhance the teleportation efficiency

In the present paper we propose a scheme for the teleportation of
a single-photon wave packet employing the EPR pair which is in the 
entangled state with respect to the "energy+time" coordinates.
EPR pairs of that kind are produced in the parametric down-conversion
processes [11]. A fundamental difference between our case and the case 
considered in Ref.[10] is that the single-photon wave packet state is 
the multi-mode one, and the parameter of time appears in the problem
explicitly.

To simplify the formulas, we shall assume that wave packet polarization 
is known. The arguments below can easily be extended to the case of 
unknown polarization by simply adding an extra subscript.
The state of a single-photon wave packet can be written as [14]
\begin{displaymath}
|1\rangle_3 = 
\int_{0}^{\infty}d\omega f(\omega)
e^{-i\omega t_0} \hat{a}^+(\omega)
|0\rangle =
\int_{0}^{\infty}d\omega f(\omega)
e^{-i\omega t_0} 
|\omega\rangle_3,
\end{displaymath}
\begin{displaymath}
[\hat{a}(\omega),\hat{a}^+(\omega')]=I\delta(\omega-\omega'),\quad
\int_{0}^{\infty} |f(\omega)|^2  d\omega =1,
\end{displaymath}
where $\hat{a}^+(\omega)$, $\hat{a}(\omega)$ are the creation and
annihilation operators of a single-mode Fock state $|\omega\rangle_3$, 
$|0\rangle$ is the vacuum state, $f(\omega)$ is the packet amplitude, 
$t_0$ is the initial moment of time which in the following will 
be assumed to be incorporated into the definition of $f(\omega)$. 
The density matrix at an arbitrary time is
\begin{equation}
\rho(3)=
\left(
\int_{0}^{\infty}d\omega e^{-i\omega t} f(\omega)|\omega\rangle_3
\right)
\left(
\int_{0}^{\infty}d\omega' {}_3\langle \omega| e^{-i\omega' t} 
f^*(\omega')
\right) 
\end{equation}
The state of the EPR-pair of photons can be written as
\begin{equation}
|1_1,1_2\rangle = 
\int_{0}^{\infty}\int_{0}^{\infty}d\omega d\omega' 
F(\omega,\omega')\hat{a}^{+}_{1}(\omega)\hat{a}^{+}_{2}(\omega')
|0\rangle =
\int_{0}^{\infty}\int_{0}^{\infty}d\omega d\omega' 
F(\omega,\omega')|\omega\rangle_1\otimes|\omega'\rangle_2,
\end{equation}
\begin{displaymath}
\rho_{_{EPR}}(1,2)=|1_1,1_2\rangle \langle 1_1,1_2|,\quad
\int_{0}^{\infty} \int_{0}^{\infty}
|F(\omega,\omega')|^2
d\omega d\omega' = 1,
\end{displaymath}
where $F(\omega,\omega')$ is the joint amplitude of the photons in the 
EPR-pair. It is important that the amplitude $F(\omega,\omega')$ 
does not factorize: ($F(\omega,\omega')\neq f(\omega)\cdot f(\omega')$). 

According to the general scheme [12,13], quantum mechanical measurements
are described by positive operators realizing the identity resolution.
Corresponding to the observables associated with the self-adjoint
operators are the orthogonal identity resolutions. Parameters
(time, rotation angles) are not related to any self-adjoint operators, 
and therefore they correspond to non-orthogonal identity resolutions
[12,13]. 

Let us first briefly discuss the joint measurements of the photons
comprising an EPR-pair. Measurement of the energy (frequency) of one of the
photons is described by the orthogonal identity resolution
\begin{equation}
\int_{0}^{\infty}E(d\Omega)=I,\quad 
E(d\Omega)=|\Omega\rangle \langle \Omega| d\Omega.
\end{equation}
Accordingly, the measurement of time is given by the non-orthogonal identity
resolution
[12,13]
\begin{equation}
\int_{-\infty}^{\infty} M(dt)=I,
\end{equation}
\begin{displaymath}
M(dt)=
\left(\int_{0}^{\infty}e^{i\Omega t}|\Omega\rangle d\Omega\right)
\left(\int_{0}^{\infty}e^{-i\Omega' t}\langle\Omega'| d\Omega'\right)
\frac{\textstyle dt}{\textstyle 2\pi}.
\end{displaymath}
Joint measurement of the energy of photons in the EPR-pair yields
the outcome probability distribution
\begin{equation}
\mbox{Pr}\{d\omega_1 d\omega_2\}=
\mbox{Tr}\{\rho_{_{EPR}}(1,2)E(d\omega_1)E(d\omega_2)\}=
|F(\omega_1,\omega_2)|^2 d\omega_1 d\omega_2,
\end{equation}
where $E(d\omega_1)$, $E(d\omega_2)$ are the projectors on the
single-mode Fock states of the first and second photon in the EPR-pair. 
Joint measurement of time (positive outcomes in the intervals $(t_1,t_1+dt_1)$ 
and $(t_2,t_2+dt_2)$) generates the probability distribution
\begin{equation}
\mbox{Pr}\{dt_1 dt_2\}=
\mbox{Tr}\{\rho_{_{EPR}}(1,2)M(dt_1)M(dt_2)\}=
|F(t_1,t_2)|^2 dt_1 dt_2,
\end{equation}
\begin{displaymath}
F(t_1,t_2)=
\frac{\textstyle 1}{\textstyle 2\pi}
\int_{0}^{\infty}\int_{0}^{\infty}
F(\omega,\omega')
e^{-i(\omega t_1 + \omega' t_2)} d\omega d\omega'.
\end{displaymath}
To illustrate the above arguments and simplify further analysis, we shall
present the joint amplitude of two correlated variables in the form
(e.g. see Ref.[15])
\begin{equation}
F(\omega,\omega')=
\frac{\textstyle 1}{\textstyle \sqrt{2\pi\sigma^2\sqrt{1-\mu^2}}}
e^{-\frac{\textstyle P(\omega,\omega')}{\textstyle 2}},
\end{equation}
where
\begin{displaymath}
P(\omega,\omega')=
\frac{\textstyle (\omega-\Omega_1)^2+(\omega'-\Omega_2)^2-
2\mu(\omega-\Omega_1)(\omega'-\Omega_2) }
{\textstyle 2\sigma^2(1-\mu^2)},
\end{displaymath}
\begin{displaymath}
\Omega_0=\Omega_1+\Omega_2,
\end{displaymath}
where $\mu$ is the correlation coefficient. The frequency $\Omega_0$ 
is assumed to be a known external parameter which is governed by
the pumping frequency in the parametric down-conversion process.

Ideal EPR correlations (anticorrelation) correspond to $\mu=-1$ ($\mu=1$). 
For $\mu\rightarrow -1$, $\sigma\rightarrow\infty$, and 
$\sigma^2(1-\mu^2)\rightarrow 0$, the joint measurement of the frequencies
of the photons belonging to the EPR pair yields the following measurement
outcomes probability distribution:
\begin{equation} 
\mbox{Pr}\{d\omega_1 d\omega_2\}\propto 
e^{ -\frac{\textstyle (\omega_1+\omega_2-\Omega_0)^2}
{\textstyle 2\sigma^2(1-\mu^2)} } 
d\omega_1 d\omega_2
\rightarrow
\delta(\omega_1+\omega_2-\Omega_0)d\omega_1 d\omega_2,
\end{equation}
and, accordingly, measurement of time
\begin{equation}
\mbox{Pr}\{dt_1 dt_2\}\propto
e^{ -\frac{\textstyle \sigma^2(t_1+t_2+2\mu t_1 t_2)^2}{\textstyle 2} } 
dt_1 dt_2 \rightarrow 
\delta(t_1-t_2)dt_1 dt_2,
\end{equation}
If the measurements are performed at spatially separated points,
the times in Eq.(10) should be understood as the reduced times corrected for
the photon times-of-flight,
$(t_{1,2}\rightarrow t_{1,2}-\frac{\textstyle x_{1,2}}{\textstyle c}$). 
For brevity, in the rest of the paper we shall always use the reduced times
without explicitly mentioning it.

Eqs. (9) and (10) represent the EPR effect for two complementary
alternatives, the dynamical energy variable and the time parameter,
similar to the EPR effect for the dynamical variables of momentum 
and position [5]. If two distant users both perform the frequency 
measurements and one of them obtained a non-zero result at a frequency
$\omega_1$, the outcome of the second measurement can be 
predicted with certainty without actually doing it: 
non-zero outcome will occur for the frequency $\omega_2=\Omega_0-\omega_1$. 
However, if the time measurements were performed and the first user 
had a non-zero outcome at the time moment $t_1$, 
the moment of photon detection by the second user is bound to be 
$t_2=t_1-\frac{\textstyle x_2-x_1}{\textstyle c}$.

The idea of teleportation applied to the present case is to
employ a joint (entangled) "energy+time" measurement of the
pair of photons one of which belongs to the EPR-pair and the second
one is in an unknown state. The indicated measurement is given by
the non-orthogonal identity resolution
\begin{equation}
\int\int M(dtd\Omega_{-})=I,
\end{equation}
\begin{displaymath}
M(dtd\Omega_{-})=
\end{displaymath}
\begin{equation}
\left(\int d\Omega_{+} e^{i\Omega_{+}t}|\Omega_{+}+\Omega_{-}\rangle_1
\otimes|\Omega_{+}-\Omega_{-}\rangle_3\right)
\left(\int d\Omega_{+}' e^{-i\Omega_{+}'t}{}_3\langle\Omega_{+}'-\Omega_{-}|
\otimes{}_1\langle\Omega_{+}'+\Omega_{-}|\right)
\frac{\textstyle dtd\Omega_{-}}{\textstyle 2\pi}
\end{equation}
Here integration is performed over the frequencies resulting in positive 
arguments of the Fock states.
It should be emphasized that the frequency
$\Omega_{-}$ is common to  all bra and ket states. 
Indeed, the integration over $t$ in Eq.(11) yields 
$\delta(\Omega_{+}-\Omega_{+}')$, and further integration over 
$\Omega_{+}'$  eliminates one integral over $\Omega_{+}'$ resulting in
\begin{equation}
\int \int d\Omega_{-} d\Omega_{+} 
\left(
|\Omega_{+}+\Omega_{-}\rangle_1\otimes|\Omega_{+}-\Omega_{-}\rangle_3
\right)
\left(
{}_3\langle\Omega_{+}-\Omega_{-}|\otimes{}_1\langle\Omega_{+}'+\Omega_{-}|
\right)=
\end{equation}
\begin{displaymath}
\int_{0}^{\infty}\int_{0}^{\infty} d\omega_1 d\omega_3 
|\omega_1\rangle_1\otimes|\omega_3\rangle_3
\mbox{ }
{}_3\langle\omega_3|\otimes{}_1\langle\omega_1|
=I_1\otimes I_3,
\end{displaymath}
where $\omega_1=\Omega_{+}+\Omega_{-}$ ? $\omega_3=\Omega_{+}-\Omega_{-}$.
In some sense the measurement (11) resembles a partial Fourier transform
over the sum frequency of the two photons (while Fourier transform over the
difference frequency is not performed).
The probability to find the measured quantities within the intervals
$(t,t+dt)$ ? $(\Omega_{-},\Omega_{-}+d\Omega_{-})$ is described by the formula
\begin{equation}
\mbox{Pr}\{dt d\Omega_{-}\}=
\left(
\int d\omega_2 \int \int d\Omega_{+}d\Omega_{+}' 
\tilde{F}(\Omega_{+}+\Omega_{-}+\omega_2;\Omega_{+}+\Omega_{-}-\omega_2)
\right.
\end{equation}
\begin{displaymath}
\left.
\tilde{F}^*(\Omega_{+}'+\Omega_{-}+\omega_2;\Omega_{+}'+\Omega_{-}-\omega_2)
f(\Omega_{+}-\Omega_{-})f^*(\Omega_{+}'-\Omega_{-})
e^{i(\Omega_{+}-\Omega_{+}')t}
\right)
\frac{\textstyle dt d\Omega_{-}}{\textstyle 2\pi}
\end{displaymath}
Here we have introduced for convenience the notation 
$\tilde{F}(\omega+\omega';\omega-\omega')\equiv F(\omega;\omega')$.
After the measurement user {\it A} sends the obtained values of $t$ and 
$\Omega_{-}$ through a public channel to user {\it B} who uses
them to reconstruct the state to be teleported.

In the limit of ideal EPR-correlations: $\sigma^2\rightarrow\infty$
and $\sigma^2(1-\mu^2)\rightarrow 0$ $(\mu\rightarrow -1)$, 
\begin{equation}
\tilde{F}(\omega+\omega';\omega-\omega')\propto 
\delta(\omega+\omega'-\Omega_0)\cdot const(\omega-\omega'),
\end{equation}
where $const(\omega-\omega')$ is a function which is almost constant
in a wide range of its argument.
After the measurement performed by user {\it A}, the state of the 
second photon in the EPR-pair observed by user {\it B} is given 
by the density matrix
\begin{equation}
\rho(2)=
\frac{\textstyle \mbox{Tr} 
\{ \rho_{_{EPR}}(1,2)\otimes\rho(3)M(dt d\Omega_{-}) \} }
{ \textstyle \mbox{Pr}\{dt d\Omega_{-}\} }
\end{equation}
In the limit of ideal EPR-correlations the state (15) observed by user {\it B}, 
taking into account Eq.(14), tends to  
\begin{equation}
\rho(2)\rightarrow
\left(
\int_{0}^{\infty}d\omega_2 e^{-i\omega_2 t}f(\omega_0-\omega_2)
|\omega_2\rangle_2 
\right)
\left(
\int_{0}^{\infty}d\omega_2' 
{}_2\langle\omega_2'| e^{i\omega_2' t}f^*(\omega_0-\omega_2')
\right),
\end{equation}
where $\omega_0=\Omega_0-\Omega_{-}$.

The density matrix (17) is almost identical to the original density matrix
to within a frequency shift in the argument  $f(\omega_0-\omega_2)$. 
The measurement similar to Eq.(12) can be used when the 
continuous dynamical variables we are dealing with are the
momentum and position. In that case it can be written in the form
\begin{displaymath}
\int\int M(dp_{-}dx_{+})=
\int_{-\infty}^{\infty} \int_{-\infty}^{\infty}
\frac{\textstyle dp_{-}dx_{+}}{\textstyle 2\pi}
\left(\int_{-\infty}^{\infty} 
dx_{-} e^{i x_{-}p_{-}}|x_{-}+x_{+}\rangle_1
\otimes|x_{-}-x_{+}\rangle_3\right)
\end{displaymath}
\begin{equation}
\left(\int_{-\infty}^{\infty} 
dx_{-}' e^{-i x_{-}'p_{-}}{}_3\langle x_{-}'-x_{+}|
\otimes{}_1\langle x_{-}'+ x_{+}|\right),
\end{equation}
??? $x_{\pm}=x_1 \pm x_2$, $x_{\pm}'=x_1' \pm x_2'$, 
$p_{+}=p_1 + p_2$.
A fundamental difference between Eqs (12) and (18) is that
the position and momentum shifts comprise a group. On the contrary,
in the case of energy and time the frequency shifts comprise a 
semigroup since frequency belongs to a semi-infinite interval
$(0,\infty)$ [12,13]. 
Therefore, in the "position+momentum" teleportation scheme
the shift in the state amplitude argument $f(x_0-x_2)$ (see Eq.(17)) 
is of no importance and can be eliminated by choosing an appropriate
reference frame
(shifting the argument $|x_2\rangle\rightarrow|x_0-x_2\rangle$; 
this shift is allowed since the argument changes within the infinite interval
$(-\infty,\infty)$).
In our opinion, in our case the arising frequency shift cannot be eliminated
by a similar modification of the argument since frequency can only take 
positive values. In this way the peculiarity of time which is a parameter
rather than a dynamical variable in quantum mechanics is manifested.

In the limit when the joint amplitude in the EPR-pair $\tilde{F}$ as
a delta-function as a function of sum frequency and a constant as a 
function of difference frequency, Eq.14 implies that the outcome 
probability distribution does not depend on time. In that case the
probability to obtain a non-zero result is the same throughout the
whole interval of $t$ $(-\infty,\infty)$. The latter means that the
ideal teleportation is necessarily associated with the process efficiency 
tending to zero, since user {\it A} receives the classical information with 
only vanishingly small probability. When a continuous variable is teleported,
the replica cannot be a perfect copy of the unknown state.

The problem of teleportation optimization when dealing with the
continuous parameters or dynamical variables still remains open.
The measurement (12) can be realized by mixing two photons
(the unknown photon and one of the photons belonging to the EPR-pair)
by a beam splitter and subsequent measurements by narrow-band
(for $\Omega_{-}$) and wide-band (for $t$) photodetectors
in the two channels.

The author is grateful to Prof. R.Laiho for discussions, and thanks Wihuri Laboratory
at the University of Turku where this study was performed for hospitality. 
This work was supported by the Russian Foundation for Fundamental Research
(grant 96-02-18918), and the Russian State Program "Advanced technologies
in micro-  and nanoelectronics".

\end{document}